\documentclass[twocolumn,showpacs,preprintnumbers,amsmath,amssymb]{revtex4}

\usepackage{graphicx}
\usepackage{dcolumn}
\usepackage{bm}

\newcommand{\eqb}{\begin{equation}}
\newcommand{\eqe}{\end{equation}}

\newcommand{\eab}{\begin{eqnarray}}
\newcommand{\eae}{\end{eqnarray}}

\newcommand{\La}{\Lambda}

\begin{document}
\title{A thermal theory for charged leptons}
\author{Ralf Hofmann}\vspace{0.3cm}
\address{Institut f\"ur Theoretische Physik,
Universit\"at Heidelberg,
Philosophenweg 16,
69120~Heidelberg,
Germany}

\begin{abstract}
We propose that charged leptons are the `baryons' of the gauge dynamics subject to a 
SU(2)$_{e}$$\times$SU(2)$_{\mu}\times$SU(2)$_{\tau}$ symmetry. 
The `mesons' of this theory are neutral and thus 
candidates for dark matter. There is, in addition, a gauge group SU(2)$_{\tiny\mbox{CMB}}$ generating the CMB photon. 
We {\sl compute} the value of $\alpha_{\tiny\mbox{em}}$ and the contribution to $\La_{\tiny\mbox{cosmo}}$ coming 
from SU(2)$_{\tiny\mbox{CMB}}$. We qualitatively explain the occurrence of dark matter, 
large intergalactic magnetic fields, 
a lepton asymmtry, the stability of $T_{\tiny\mbox{CMB}}$, 
and the deviation from thermal QED behavior of 
hot plasmas present in the interior of the sun and 
terrestial nuclear fusion experiments. 
 
\end{abstract}

\pacs{12.38.Mh,11.10.Wx,12.38.G,04.40.Nr}

\maketitle

\indent 
In \cite{PRL1} we have developed an analytic, inductive approach to hot SU(N) 
pure gauge dynamics. Here we apply this approach. 
In the approach \cite{PRL1} we have observed that an SU(2) gauge theory develops 
an electric phase (2$^{\mbox{\tiny{nd}}}$ order thermal phase transition), presumably at a temperature $T^P_E$ 
tremendously higher than $T_c$ (deconfinement temperature), maybe we have 
$T^P_E\sim M_P$ ($M_P$ the Planck mass). In the electric phase SU(2) is spontaneously broken 
to U(1) by an adjoint Higgs field $\phi$ describing the 
condensation of calorons \cite{PRL1}. The tree-level massless (TLM) gauge boson - the photon - 
acquires mass by a one-loop radiative tadpole correction, see Fig. 1. 
\begin{figure}
\begin{center}
\leavevmode
\leavevmode
\vspace{6.5cm}
\includegraphics{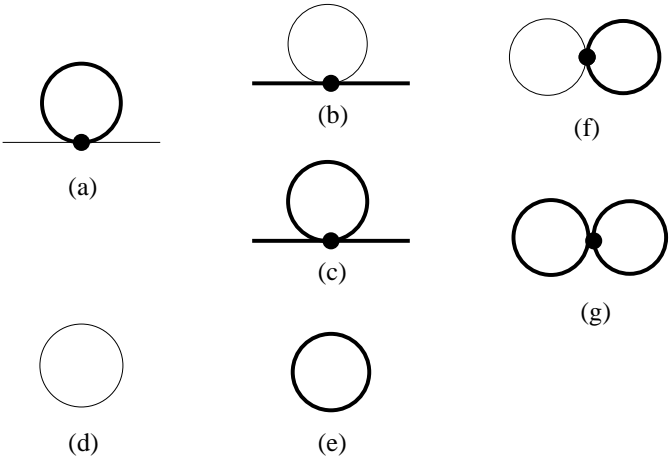}
\end{center}
\caption{Radiative corrections arising in the electric phase 
if gauge-boson fluctuations are assumed to be
noninteracting thermal quasiparticles. 
(a) - (c) thermal mass corrections and (d) - (g) 
corrections to the vacuum pressure. 
Thin lines are associated with tree-level massless (TLM) modes and 
thick ones with tree-level heavy (TLH) modes.}      
\end{figure} 
This mass 
is given as 
\eqb
\label{massTLM}
m_{\tiny\mbox{TLM}}=0.433/(\sqrt{8}\pi)\sqrt{\La_E^3/(2\pi T)}+O(1/e^2)\,,
\eqe
where $\La_E$ denotes the mass scale governing 
the electric phase, and $e$ is the (large) SU(2) gauge coupling. As a function of $T$ 
$m_{\tiny\mbox{TLM}}$ has a global maximum at the critical temperature $T_M^c$ where a 
mild 1$^{\mbox{\tiny{st}}}$ order thermal phase 
transition towards the magnetic phase takes place. 
The mass $m_{\tiny\mbox{TLM}}$ decreases with $T^{-1/2}$ as $T$ grows. Taking $T$ to be room 
temperature, $T\sim 6.5\times 10^{-2}$\,eV, the temperature of visible radiation, 
$T\sim 6.5$\,eV, the temperature in the interior of the sun, 
$T\sim 10^{7}$\,eV and assuming $\La_E\sim 3.91\times 10^{-4}$\,eV (see below) we arrive 
at a thermal photon mass in the electric phase of about 
$5.9\times 10^{-7}$\,eV, $5.9\times 10^{-8}$\,eV, 
$4.8\times 10^{-11}$\,eV, respectively. The {\sl thermal} photon mass is not to be confused with the only 
valid bound on the photon mass ($\sim 10^{-14}$\,eV) coming from precision measurements 
of the Coulomb potential \cite{Dvali2003,Williams1971}, see below. 
In unitary gauge off-diagonal gauge bosons ($V^\pm$) have large tree-level masses (TLH mode \cite{PRL1}) 
with small radiative corrections throughout the entire electric phase. 
There is {\sl one} species of 
quasiclassical, stable BPS magnetic monopoles 
and antimonopoles in the electric phase \cite{Nahm1984,KraanVanBaal1998,PRL1} 
to which the massive gauge bosons ($V^\pm$) do not couple.

Under the intact U(1) gauge symmetry in the electric phase the monopole charge 
$g$ is given as $g=2\pi/e$ \cite{JackiwRebbi1976,Hasenfratz'tHooft}. As the system cools 
down the electric coupling $e$ strongly rises from a 
value very close to zero at $T^P_E$ and then quickly 
relaxes to a value $e\sim 17.15$. Close to a temperature $T_M^c\ll T^P_E$ a regime of logarithmic blow-up is 
reached. At $T_M^c$ $V^\pm$ gauge bosons decouple kinematically. 

The value $e\sim 17.15$, see Fig. 2, is {\sl independent} of the 
initial condition $e(T^P_E)=0$ as long as $T^P_E$ is 
sufficiently larger than $T_M^c$. Once the system has undergone the transition to the 
magnetic phase by condensing monopoles, the left-over photon is {\sl exactly} massless \cite{PRL1} on the magnetic 
side of the phase boundary. In the magnetic 
phase the gauge coupling $g$ and the photon mass quickly increases 
with decreasing temperature, see Fig. 1. The ground state of the system becomes 
{\sl superconducting} with respect to the 
spontaneously broken U(1) gauge symmetry. Considerably far away from the transition 
center vortices form due to a growing size of monopoles in the condensate \cite{PRL1}. 
In the magnetic phase, single crossings of center 
vortices (topological charge 1/2) have charge $2\pi/g$ with respect to the massive photon 
\cite{ReinhardtSchroderTokZhukovsky2002}. At the critical temperature 
$T^c_M<T^c_E$ the magnetic gauge coupling $g$ diverges logarithmically: photons 
decouple kinematically, and the 
system condenses the center vortices. A 1$^{\tiny{\mbox{st}}}$ order nonthermal phase 
transition takes place \cite{PRL1}. There 
are stable `hadrons' with what we call one unit of electric charge 
in the Standard Model: `baryons'. Moreover, 
there is a stable neutral `hadron', see Fig. 3. 
Considering {\sl naively} only a single SU(2) gauge group, 
the value of the electric charge is $g=2\pi/e\sim 0.3664$.
\begin{figure}
\begin{center}
\leavevmode
\leavevmode
\vspace{3.5cm}
\includegraphics{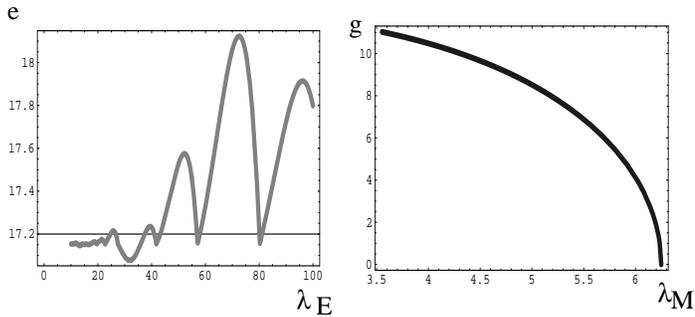}
\end{center}
\caption{The gauge couplings in the electric and in the magnetic phase as a 
function of dimensionless temperature $\lambda_{E,M}$, see \cite{PRL1}. The `nervous' behavior in 
the electric case is due to numerical interpolation errors. 
The plateau value of $e$ can be read off at the downward peaks. 
It is $e=17.15$.}      
\end{figure} 
This would yield $\alpha_{\tiny\mbox{em}}=g^2/(4\pi)\sim 1/93.6$: 
a value which is considerably off 
the measured value $\alpha_{\tiny\mbox{em}}\sim 1/137$. So what went wrong in 
our naive approach to the electric charge of a lepton? 

We propose that the CMB photon and the spectrum of charged leptons 
is described by 
gauge dynamics subject to the symmetry 
SU(2)$_{\tiny\mbox{CMB}}\times$SU(2)$_{e}$$\times$SU(2)$_{\mu}\times$SU(2)$_{\tau}$ 
(CMB stands for cosmic microwave background). The choice of SU(2) factors derives 
from the experimental fact that no mass-degenerate 
charged leptons exist in nature, see below. The scale of the first factor is set by the CMB 
temperature $T_{\tiny\mbox{CMB}}=2.728$\,K$\sim 2.182\times 10^{-4}$\,eV, 
the scales of the other factors are set by the respective 
masses of the charged leptons. The gauge dynamics subject to the 
SU(2)$_{e}$$\times$SU(2)$_{\mu}\times$SU(2)$_{\tau}$ is in the confining phase: 
no asymptotic gauge bosons exist here and the surviving, effective symmetry is a spontaneously 
broken {\sl local} Z$^{e}_2$$\times$Z$^{\mu}_2\times$Z$^{\tau}_2$ 
symmetry \cite{PRL1}. Charged leptons are the `baryons' of 
these gauge theories, `mesons' are neutral and much less massive, see Fig. 3. 
The `mesons' are candidates for dark matter, an interpretation of 
these particles is given in \cite{Hofmann20041}. In \cite{PRL3} 
we have shown that a local Z$_2$ symmetry favors 
Fermi-Dirac statistic over Bose-Einstein statistics by a factor two if we assume 
the thermalization of the respective stable `hadronic' states. In particular, the `baryons' 
electron, muon, and tau are rendered fermions as a result of 
genuinely local Z$_2$ transformations. 

The dynamics subject to SU(2)$_{\tiny\mbox{CMB}}$ is 
nonconfining. This theory is in its magnetic phase close to the electric phase 
boundary, where the photon is almost massless, see below and Fig.\,4. 
At the temperature $T_{\tiny\mbox{CMB}}$ the CMB 
photon is {\sl quantum entangled} with the (infinitely) heavy photons 
of the gauge theory with SU(2)$_{e}$$\times$SU(2)$_{\mu}\times$SU(2)$_{\tau}$ symmetry due 
to their mixing during an early epoch in the evolution of the Universe 
where the temperature was higher than the muon but lower than the pion mass. 

Using this input, we now {\sl derive} the value of $\alpha_{\tiny\mbox{em}}$. Since the $\tau$ lepton has a 
very small life-time ($\delta t_\tau/\delta t_\mu\sim 10^{-7}$ but $m_\mu/m_\tau=6\times 10^{-2}$, decay by 
electroweak interactions) the photons of $SU(2)_\tau$ scattered by this 
particle were most probably never thermalized. Moreover, the gauge dynamics due to 
$SU(2)_\tau$ is contaminated by strong interactions for temperatures larger than the pion mass. But 
we only want to consider pure electromagnetism here. At $m_\mu<T<m_\pi$ we thus consider the gauge 
group SU(2)$_{\tiny\mbox{CMB}}\times$SU(2)$_{e}$$\times$SU(2)$_{\mu}$ only. We then have three species 
of thermalized photons $a^i_\mu$ ($i=\mbox{CMB},e,\mu$) in the plasma which mix quantum mechanically 
by a thermal mass matrix (recall, that the tree-level massless modes 
acquire a mass by radiative corrections in the electric phase, Fig. 1 and (\ref{massTLM})). 
In a thermal state it is very reasonable to assume 
that photon wave functions $\tilde{a}_\mu^i$ are generated by 
maximal mixing of the photon states originating from the respective 
dynamics of each SU(2) factor, for example
\eab
\label{ecs}
\tilde{a}_\mu^{\tiny\mbox{CMB}}&=&1/\sqrt{3}(a_\mu^{\tiny\mbox{CMB}}+a^{e}_\mu+a^{\mu}_\mu)\,.
\eae
and similar for the orthogonal superpositions $\tilde{a}_\mu^{e}$ and $\tilde{a}_\mu^{\mu}$. 
When the temperature falls below $m_\mu$, an epoch of exponential cosmological 
inflation of scale $m_\mu$, which is terminated by a nonthermal 1$^{\tiny\mbox{st}}$ order transition, 
renders the $SU(2)_{\mu}$ gauge theory confining. The photon $a^{\mu}_\mu$ decouples kinematically and 
is almost instantaneously removed from (\ref{ecs}) as a fluctuating degree of
freedom. This changes the wave function normalization 
of the remaining two photons by a factor $\sqrt{2}/\sqrt{3}$. The new maximal-mixing superposition still 
sees the charge of the muon because of a quantum entanglement with the kinematically 
decoupled photon $a^{\mu}_\mu$. The coherent superposition of $a_\mu^{\tiny\mbox{CMB}}$ 
and $a^{e}_\mu$ is now normalized as
\eab
\label{ecsr}
\tilde{a}_\mu^{CMB}&=&1/\sqrt{2}\left[\sqrt{2}/\sqrt{3}(a_\mu^{\tiny\mbox{CMB}}+a^{e}_\mu)\right]\,.
\eae
The further removal of $a^{e}_\mu$ from the thermal spectrum by a period of exponential inflation of scale 
$T\sim m_e$ does not imply a new normalization since only the CMB photon survives kinematically. 
In the electric phase and away from the boundary to the magnetic phase monopoles are much 
heavier than the photons, $m_{\tiny\mbox{mon}}/m_{\tiny\mbox{photon}}=64\pi^2/(\sqrt{2}e)=25.5$. 
Shortly before the transition to 
the magnetic phase they experience a strong drop in their mass and condense subsequently 
in the magnetic phase. This condensation 
slowly rotates a superposition 
of monopoles belonging to different SU(2) factors to the monopole of the 
SU(2) factor which undergoes the electric-magnetic transition (monopole excitations, which disappear in the 
magnetic phase, re-appear as charged leptons in the center phase). We conclude, that the above argument for 
photons does not apply to monopoles. The factor $\sqrt{2}/\sqrt{3}$ in (\ref{ecs}) 
reduces the wave function of the original photon or, equivalently, the 
gauge coupling $g\to g_r=\sqrt{2}/\sqrt{3}\, g$. As a consequence, the real value of $\alpha_{em}$ is
\eab
\label{alphaem}
\alpha_{\tiny\mbox{em}}&=&g_r^2/(4\pi)\sim 1/140.433\,.
\eae
The deviation from the measured value $\alpha_{em}\sim 1/137$ is 2.4\%! We attribute this small 
mismatch to the omission of radiative corrections in our evolution of 
the gauge coupling $e$ in the electric phase (recall that there is a {\sl zeroth-order} 
in $1/e^2$ radiative correction to the photon mass). The result (\ref{alphaem}) 
is an impressive experimental confirmation for the validity of the 
analytic approach to thermal SU(N) Yang-Mills theory in \cite{PRL1}. The reader may wonder why we 
could predict the value of $\alpha_{\tiny\mbox{em}}$ from an analysis in the {\sl electric} phase 
while charged leptons, `baryons' of 
SU(2)$_{e}$$\times$SU(2)$_{\mu}$$\times$SU(2)$_{\tau}$, 
are solitons in the {\sl center} phase of this theory. The solution to this apparent puzzle is 
indicated in Fig. 3 and explained in its caption. The electric charge is quantized by the single 
monopole residing in a `baryon' (or charged lepton) of spin 1/2 as the crossing point of two center vortices. 
This `baryon'  has precisely the same charge as a monopole in the electric phase 
due to the topological equivalence of a center-vortex 
crosssing and and an isolated monopole \footnote{The continuous variation of the magnetic charge in dependence 
of temperature in the magnetic phase is due to the occurrence of a monopole-antimonopole {\sl condensate}, 
that is, the percolation of single (anti)monopole worldlines.}.    

In the magnetic phase monopoles condense, 
the magnetic coupling $g$ is precisely zero at the 
magnetic-electric transition and the electric coupling $e$ is infinite. 
The critical temperature $T_E^c$ is the only point where a magnetic description 
is precisely dual to an electric description of the underlying SU(2) gauge theory. 
In the course of the magnetic-center transition, when center vortices condense, monopoles 
re-appear on the crossings of center vortices. In a given spatial region, charged 
lepton number is not conserved. This leads us to a new mechanism for the generation of lepton asymmetry. 
Recall Sakharov's conditions \cite{Sakharov1967} for this:
(i) lepton number violation, (ii) violation of thermal equilibrium, and (iii) a CP violation 
in the interaction. We just discussed the violation of electric charge (in the sense of the Standard Model) 
during the magnetic-center transition. 
So condition (i) is met. The transition to the center phase is truely 1$^{\tiny\mbox{st}}$ order. 
It is preceeded by an epoch of exponential inflation (a photon decouples kinematically 
shortly before the transition and thus the energy density $\rho_{\tiny\mbox{vac}}$ 
is the one of the ground state, the equation of state is $P_{\tiny\mbox{vac}}=-\rho_{\tiny\mbox{vac}}$). So 
condition (ii) is met. We know that weak interactions violate CP due to a $3\times 3$ CKM 
matrix for the three visible charged leptons. So there 
is a CP violating phase, condition (iii) is 
also met. The magnitude of lepton asymmetry depends on the mass of the charged 
lepton since this sets the scale for the strength of the 1$^{\tiny\mbox{st}}$ 
order transition to the center phase. It also depends on the structure of the CKM matrix. 
Lepton asymmetry should then be computable from the measured CKM 
matrix in the Standard Model and the possibility to estimate the strength of the 1$^{\tiny\mbox{st}}$ 
order transition. 
\begin{figure}
\begin{center}
\leavevmode
\leavevmode
\vspace{3.5cm}
\includegraphics{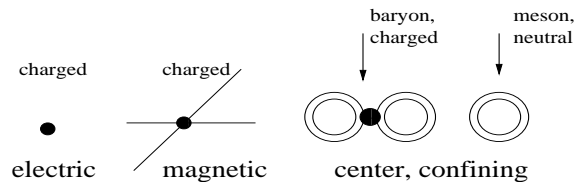}
\end{center}
\caption{The evolution of an electric charge (in the sense of the Standard Model) 
through the three phases of an SU(2) gauge theory. 
The dots indicate monopoles. The thin (double) lines in the magnetic (center) phase denote 
(fat) center vortices. The `meson' in the center phase is not 
charged under $U(1)_{\tiny\mbox{em}}$. It is a dark-matter candidate. 
In the presence of the asymptotic CMB photon there are no other stable `baryonic' states 
(more than one monopole or antimonopole in a `baryon' would either repel or annihilate).}      
\end{figure}
Let us now derive a boundary condition for the gauge theory SU(2)$_{\tiny\mbox{CMB}}$ 
from experimental information. On cosmological scales we do not see neither center vortices nor do we see 
light-mass electric charges at any temperature in laboratory 
experiments \cite{chpar} (monopoles and antimonopoles 
are extremly dilute in the electric phase). We do, however, see a massless photon 
in the CMB radiation which has an (almost) perfect thermal spectrum. This leads us to the conclusion 
that the CMB temperature $T_{\tiny\mbox{CMB}}$ is slightly lower than the critical temperature $T^c_E$ 
of the electric-magnetic transition. Only for $T_{\tiny\mbox{CMB}}+0=T^c_E$ do we have a {\sl 
condensate} of monopoles (no explicit monopoles!) and not yet an explicit 
center vortex in our theory. Moreover, in our theory the CMB photon is exactly 
massless at $T_{\tiny\mbox{CMB}}=T^c_E-0$ 
due to a logarithmic blow-up of 
$e$ on the electric side of the phase boundary rendering $g=0$. As a consequence, 
the Abelian Higgs mechanism is not operative close to the phase boundary. So the boundary condition is: 
$T_{\tiny\mbox{CMB}}=T^c_E$.

The cooling of the ground state into superconductivity 
for $T<T_{\tiny\mbox{CMB}}$ will change our universe drastically. 
The onset of this regime is already seen in the existence of 
{\sl large intergalactic magnetic fields}. The generation of these magnetic fields is driven 
by one-sided temperature fluctuations which drive the associated, intergalactic region away from the 
electric-magnetic phase boundary into the magnetic phase, see Fig.\,4. What is the mechanism that 
prevents us from running deep into the superconducting regime? The thermal mass of the CMB photon in the electric 
phase jumps from $m_{\tiny\mbox{TLM}}=0.433/(\sqrt{8}\pi)\sqrt{\La_E^3/(2\pi T_{\tiny\mbox{CMB}})}
\sim 10^{-5}$\,eV to zero across the phase boundary. 
In the magnetic phase the CMB photon acquires mass as temperature drops. 
As a function of temperature this creates a potential for the photon mass 
with a wall of height $\sim 10^{-5}$\,eV to the right and smoothly increasing mass 
to the left of the minimum at $T=T_{\tiny\mbox{CMB}}$, see Fig. 4. 
\begin{figure}
\begin{center}
\leavevmode
\leavevmode
\vspace{3.5cm}
\includegraphics{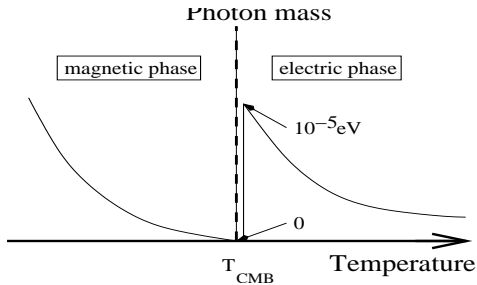}
\end{center}
\caption{Photon-mass dependence on temperature.}      
\end{figure}
This effect keeps the CMB temperature stable for about another present 
age of the universe. The energy needed for gravitational expansion is taken from the ground state. 
The stabilizing effect is actually seen in a lattice simulation of the energy 
density $\rho$ as a small dip of $\rho$ shortly before the deconfinement transition 
\cite{EKSM1982}. In the far-distant future we will have to cope with increasing 
magnetic fields and magnetic vortices in our universe until a phase of exponential 
inflation will put to an end the existence of the CMB photon.  
 
Since the masses of charged leptons are 
much higher than $T_{\tiny\mbox{CMB}}$ ($m_e=5\times 10^5\,\mbox{eV}\sim 2.3\times 10^9\,T_{\tiny\mbox{CMB}}\sim \La_e$!) 
the dynamics of SU(2)$_e$, SU(2)$_\mu$, and SU(2)$_\tau$ 
are all in their confining phase at $T_{\tiny\mbox{CMB}}$. The vacuum energy densities and the vacuum pressures 
in all three theories are {\sl precisely} zero \cite{PRL1} (this is protected by local Z$_2$ symmetries). 
Let us now calculate the electromagnetic, homogeneous contribution to the 
cosmological constant. At the electric-magnetic phase boundary we have 
$\lambda_M(0)=(1/4)^{1/3}\times 9.193$ where due to our above boundary condition 
$\lambda_M(0)\equiv 2\pi T_{\tiny\mbox{CMB}}/\La_M$ and $\La_M$ denotes the magnetic scale 
of SU(2)$_{\tiny\mbox{CMB}}$, see \cite{PRL1}. Prescribing $T=T_{\tiny\mbox{CMB}}=2.728\,K=2.1824\times 10^{-4}\,$eV, 
yields $\La_M=2.368\times 10^{-4}\,$eV. The resulting vacuum energy density is 
$1/2\,V_M=\pi T_{\tiny\mbox{CMB}}\La_M^3=(3.1\times10^{-4}\,\mbox{eV})^4$. This is smaller than the commonly 
accepted value $\La_{\tiny\mbox{cosmo}}\sim (10^{-3}\,\mbox{eV})^4$. There are two possible reasons for this. 
First, we have used a tree-level analysis to determine the critical value $T^c_E$ in our approach. Including radiative 
corrections in the running of the electric gauge coupling $e$ should yield a slightly 
smaller value than $\lambda_E^c=9.193$ since massive TLM 
modes slow down the evolution, see \cite{PRL1}. This 
should imply a slightly higher value of 
$1/2\,V_M$. Second, there are local regions 
with negative pressure induced by high-energy 
particle collisions subject to SU(N) gauge dynamics. 
After coarse-graining this effect also 
contributes to $\La_{\tiny\mbox{cosmo}}$.
 
Let us now discuss the validity of thermal QED. 
The additional photon belonging to $SU(2)_e$ can not be seen 
in terrestial experiments with hot plasmas up to $T\sim m_e=5\times10^5$\,eV. 
The highest temperatures reached in nuclear fusion experiments are 
about $T\sim 10^{7}$\,eV - the temperature in the interior of the sun. Due to the 
occurrence of a second photon, the strong dilution of the electron density in the electric phase of 
$SU(2)_{e}$, and a considerable contribution of vacuum energy 
these plasmas should show a strong deviation from standard thermal QED behavior. All solar models 
rely on standard thermal QED and thus should be revised. This should have a major impact 
on the prediction of neutrino fluxes. As for Tokamaks we predict a strong increase of the magnetic field at 
temperatures close to the electron mass and a disappearence of electronic electric charge 
at about 1.8 times this temperature. This effect could be measured by applying a large and 
homogeneous electric field to the 
Tokamak. Positively charged ions should then move to one end and no negative charge should 
be measured at the other end of the Tokamak. 

To summarize, we have proposed a strongly interacting gauge theory underlying 
electrodynamics, and we have explored some of the 
consequences of this theory. The apparent 
structurelessness of the electron and the muon, as it is seen in high-energy 
experiments for the modulus of momentum transfers 
away from the electron and muon mass, 
is explained in \cite{Hofmann20041}.

{\sl Acknowledgments:} 
It is a pleasure to thank Qaisar Shafi and Zurab Tavartkiladze for 
very useful discussions and Pierre van Baal for a 
correspondence. The hospitality and financial support of CERN's theory division are thankfully acknowledged.

\bibliographystyle{prsty}

\begin{thebibliography}{10}

\bibitem{PRL1}
R. Hofmann, hep-ph/0312046.

\bibitem{Dvali2003}
E. Adelberger, G. Dvali, and A. Gruzinov, hep-ph/0306245.

\bibitem{Williams1971}
E. R. Williams, J. E. Faller, and H. A. Hill, Phys. Rev. Lett. {\bf 26}, 721 (1971). 

\bibitem{Nahm1984}
W. Nahm, Lect. Notes in Physics. 201, eds. G. Denaro, e.a. (1084) p. 189.

\bibitem{KraanVanBaal1998}
T. C. Kraan and P. van Baal, Nucl. Phys. B {\bf 533}, 627 (1998); 
T. C. Kraan and P. van Baal, Phys. Lett. B {\bf 435}, 389 (1998). 

\bibitem{JackiwRebbi1976}
R. Jackiw and C. Rebbi, Phys. Rev. D {\bf 13}, 3398 (1976). 

\bibitem{Hasenfratz'tHooft}
P. Hasenfratz and G. 't Hooft, Phys. Rev. Lett. {\bf 36}, 1119 (1976).

\bibitem{Hofmann20041}
R. Hofmann, hep-ph/0401017. 

\bibitem{PRL3}
R. Hofmann, hep-ph/0312051. 

\bibitem{ReinhardtSchroderTokZhukovsky2002}
H. Reinhardt, O. Schroder, T. Tok, and V. Ch. Zhukovsky, Phys. Rev. D {\bf 66}, 085004 (2002). 

\bibitem{Sakharov1967}
A. D. Sakharov, Pisma Zh. Eksp. Teor.Fiz. {\bf 5}, 32 (1967); JETP Lett. {\bf 5}, 24 (1967); 
Sov. Phys. Usp. {\bf 34}, 392 (1991); Usp. Fiz. Nauk {\bf 161}, 61 (1991).

\bibitem{chpar}
This charged particle can not be seen in terrestial experiments because, locally, thermal equilibirum is always 
violated on scales much larger than $T_{\tiny\mbox{CMB}}$. The temperature of liquid $^3$He may be 
lowered to milli-Kelvins but this temperature is the temperature of heavy fermionic compounds with 
eV electrons residing in their interiors. I thank Paul Clegg for a discussion of this point. 

\bibitem{EKSM1982}
J. Engels {\sl et al.}, Nucl. Phys. B {\bf 205}[FS5], 545  (1982).

\end{thebibliography}

\end{document}